 \documentclass[12pt,preprint]{aastex}

\shorttitle{Enstatite-rich Warm Debris Dust around HD165014}
\shortauthors{Fujiwara et al.}

\begin{document}

\title{Enstatite-rich Warm Debris Dust around HD165014}


\author{
Hideaki~Fujiwara\altaffilmark{1}\footnote{present address: Institute of Space and Astronautical Science, Japan Aerospace Exploration Agency,
3-1-1 Yoshinodai, Chuo-ku, Sagamihara, Kanagawa 252-5210, Japan}, 
Takashi~Onaka\altaffilmark{1}, 
Daisuke~Ishihara\altaffilmark{2}, 
Takuya~Yamashita\altaffilmark{3}, 
Misato~Fukagawa\altaffilmark{4}, 
Takao~Nakagawa\altaffilmark{5}, 
Hirokazu~Kataza\altaffilmark{5}, 
Takafumi~Ootsubo\altaffilmark{6}, 
and Hiroshi~Murakami\altaffilmark{5} 
}


\altaffiltext{1}{Department of Astronomy, School of Science, University
of Tokyo, Bunkyo-ku, Tokyo 113-0033, Japan; hideaki@ir.isas.jaxa.jp}
\altaffiltext{2}{Graduate School of Science, Nagoya University, Furo-cho, Chikusa-ku, Nagoya 464-8602, Japan}
\altaffiltext{3}{National Astronomical Observatory of Japan, 2-21-1 Osawa, Mitaka, 
Tokyo 181-0015, Japan}
\altaffiltext{4}{Graduate School of Science, Osaka University, 1-1 Machikaneyama, 
Toyonaka 560-0043, Osaka, Japan}
\altaffiltext{5}{Institute of Space and Astronautical Science, Japan Aerospace Exploration Agency,
3-1-1 Yoshinodai, Chuo-ku, Sagamihara, Kanagawa 252-5210, Japan}
\altaffiltext{6}{Astronomical Institute, Graduate School of Science,
Tohoku university, Aramaki, Aoba-ku, Sendai, 980-8578, Japan}


\begin{abstract}
We present the {\it Spitzer}/Infrared Spectrograph spectrum of the main-sequence star HD165014, 
which is a warm ($\gtrsim 200$~K) debris disk candidate discovered by the {\it AKARI} All-Sky Survey. 
The star possesses extremely large excess emission at wavelengths longer than 5$~\micron$. 
The detected flux densities at 10 and 20$~\micron$ are $\sim 10$ and $\sim 30$ times 
larger than the predicted photospheric emission, respectively. 
The excess emission is attributable to the presence of circumstellar warm dust. 
The dust temperature is estimated as 300--750~K, corresponding to 
the distance of 0.7--4.4~AU from the central star. 
Significant fine-structured features are seen in the spectrum 
and the peak positions are in good agreement with those of crystalline enstatite. 
Features of crystalline forsterite are not significantly seen. 
HD165014 is the first debris disk sample 
that has enstatite as a dominant form of crystalline silicate rather than forsterite. 
Possible formation of enstatite dust from differentiated parent bodies 
is suggested according to the solar system analog. 
The detection of an enstatite-rich debris disk in the current study  
suggests the presence of large bodies and a variety of silicate dust processing 
in warm debris disks.
\end{abstract}




\section{Introduction}

Debris disks were discovered in main-sequence stars 
by infrared excess over the photospheric emission 
in observations by {\it IRAS} in the 1980s. 
Since debris disks are thought to be ``extra-solar zodiacal light'' 
formed via dust production from asteroids and/or comets \citep[e.g.][]{backman93,lecavelier96}, 
it is interesting to examine mineralogical characteristics of debris dust 
and to explore connection between the debris dust and the dust in the solar system. 
Debris disks are also important as a probe of planetesimals, 
building blocks of planets, in extra-solar systems. 

Mid-infrared (MIR) spectroscopy is a strong tool to investigate the properties of 
debris dust. 
Recent MIR observations with InfraRed Spectrograph \citep[IRS;][]{houck04} 
on board {\it Spitzer} \citep{werner04} 
have revealed the presence of abundant crystalline forsterite (Mg$_2$SiO$_4$) 
and silica (SiO$_2$) in several debris disks \citep[e.g.][]{chen06,lisse09}. 

In this Letter, 
we present an MIR low resolution spectrum of the main-sequence star HD165014 
obtained with {\it Spitzer}/IRS. 
HD165014 is a debris disk candidate with large MIR excess 
selected from a search for debris disks \citep{fujiwara09c}
in the {\it AKARI}/Infrared Camera (IRC) All-Sky Survey data \citep{ishihara09}.
We report the detection of abundant crystalline enstatite (MgSiO$_3$) 
dust compared to forsterite 
toward the star and discuss the origin of the debris dust around HD165014.

\section{Observations and Data Analysis}

\subsection{{\it AKARI}/IRC all-sky survey}

The S9W (9$~\micron$) and L18W (18$~\micron$) images of HD165014 were taken with the IRC \citep{onaka07} 
on board {\it AKARI} \citep{murakami07} as part of the All-Sky Survey observations from 2006 May to 2007 August \citep{ishihara09}. 
The 5$\sigma$ sensitivity for a point source per scan is estimated to be 50~mJy in the S9W band and 90~mJy in the L18W band, 
and the typical absolute uncertainty in flux density is $\sim 3$\% for the S9W band and $\sim 4$\% for the L18W band at present. 
HD165014 was observed twice in the survey with the interval of 6 months 
and the fluxes at the two periods agree with each other within the uncertainty, 
indicating no significant variations in the flux. 
The position accuracy and spatial resolution is estimated to be better than $2\arcsec$ and $10\arcsec$, respectively, 
securely concluding that the {\it AKARI} source is associated with HD165014.
HD165014 was selected as a candidate of warm debris disk with large 18$~\micron$ excess 
by a search in the {\it AKARI}/IRC All-Sky Survey data \citep{fujiwara09c}.

\subsection{{\it Spitzer}/IRS Observations}

IRS observations of HD165014 were made on 2008 November 5 (AOR ID 26122752). 
All the four low-resolution modules, Short-Low 2 (SL2; 5.2-7.7$~\micron$) and Short-Low 1 (SL1; 7.4-14.5$~\micron$), 
Long-Low 2 (LL2; 14.0-21.3$~\micron$) and Long-Low 1 (LL1; 19.5-38.0$~\micron$), were used 
to obtain a full $5-35~\micron$ low-resolution ($\lambda/\Delta \lambda \sim 100$) spectrum. 
We use the pipeline-processed (S18.1) Basic Calibrated Data products from the {\it Spitzer} Science Center 
and analyzed the sky-subtracted extracted one-dimensional spectral data ``bksub.tbl'' 
since the target is a point source in a relatively empty field. The wavelength calibration is as good as 0.1$~\micron$ along the dispersal direction. 
For the LL1 spectra, we use the data only for $\lambda < 35~\micron$ because the noise becomes large at $\lambda > 35~\micron$, 
whose spectral range is not crucial for the present analysis. 
The absolute flux accuracy in the SL and LL spectra of the pipeline-processed products is better than 10\%.

\subsection{Subaru/COMICS Observations}

HD165014 was observed with the COoled Mid-Infrared Camera
and Spectrometer \citep[COMICS;][]{kataza00} 
mounted on the 8 m Subaru Telescope on 2008 July 16 and 17. 
Imaging observations in the $8.8~\micron$, $11.7~\micron$, and $18.8~\micron$ bands were carried out. 
The secondary mirror chopping method was used for the background cancellation.
We used standard stars ($\gamma$~Aql and $\epsilon$~Sco) from \citet{cohen99} as a flux calibrator 
and the reference point-spread functions were derived from the observations.
We observed the standard stars before or after the observations of HD165014 in the same manner as HD165014. 
For the data reduction, we used our own reduction tools and IRAF \citep{tody93}. 
The standard chop pair subtraction and the shift-and-add method 
in units of 0.1 pixel were employed. 
We applied airmass correction using ATRAN \citep{lord92}.
The derived flux densities are summarized in Table~\ref{photometry}.

\subsection{Photospheric Fitting}

The photospheric flux density of HD165014 is estimated from the Kurucz model \citep{kurucz92}
fitted to the $BVI$- and Two Micron All Sky Survey \citep{skrutskie06} $JHK_{\rm S}$-band photometry of the star taking account of the extinction. 
The only available spectral classification of HD165014 (F2V) is provided by the Michigan Catalog \citep{houk75}. 
We reexamine the spectral type of the star by the photospheric fitting. 
We use the Kurucz models of various spectral-type dwarfs 
for the stellar photospheric template of HD165014 
and the distance $d$ and the extinction $A_V$ are set free. 
We adopt an extinction curve $A_\lambda/A_V$ given by \cite{fitzpatrick09}, 
\begin{eqnarray}
\frac{A_\lambda}{A_V}= \left( \frac{0.349 + 2.087R_V}{1+(\lambda/0.507)^{2.05}}-R_V \right ) \frac{1}{R_V}, 
\end{eqnarray}
where $A_\lambda$ is the extinction toward the star at a wavelength $\lambda$ (in $\micron$) 
and the ratio of the total to the selective extinction $R_V$ is assumed as $3.11$. 

The photospheric fitting shows that  
the $BVIJHK_{\rm S}$-band photometry of HD165014 is well accounted for with almost the same significance
by spectral types between B8V ($T_{\rm eff}=12000$~K) with $A_V=3.0$ at $d=190$~pc 
and F2V ($T_{\rm eff}=7000$~K) with $A_V=1.9$ at $d=70$~pc. 
It is difficult to determine the spectral type of the star more accurately only from photometric data due to its large extinction 
and optical high-resolution spectroscopy is required for further examination. 
An earlier spectral type at a larger distance seems more reasonable for HD165014, taking account of the large amount of extinction. 
In the following we assume A0V (with $A_V=2.7$ and $d=140$~pc) as the spectral type of HD165014 rather than F2V given by the Michigan Catalog. 
The estimated extinction $A_V=2.7$ is still large for a star at 140~pc, suggesting that part of extinction may be of circumstellar origin.
The derived photospheric flux density from the fitting in each MIR band, 
which does not depend on the spectral type sensitively (only a few percent of variation from B8V to F2V), 
is listed in Table~\ref{photometry}.

\section{Results}

\subsection{Spectral Energy Distribution}

\begin{figure}
\epsscale{1.1}
\plotone{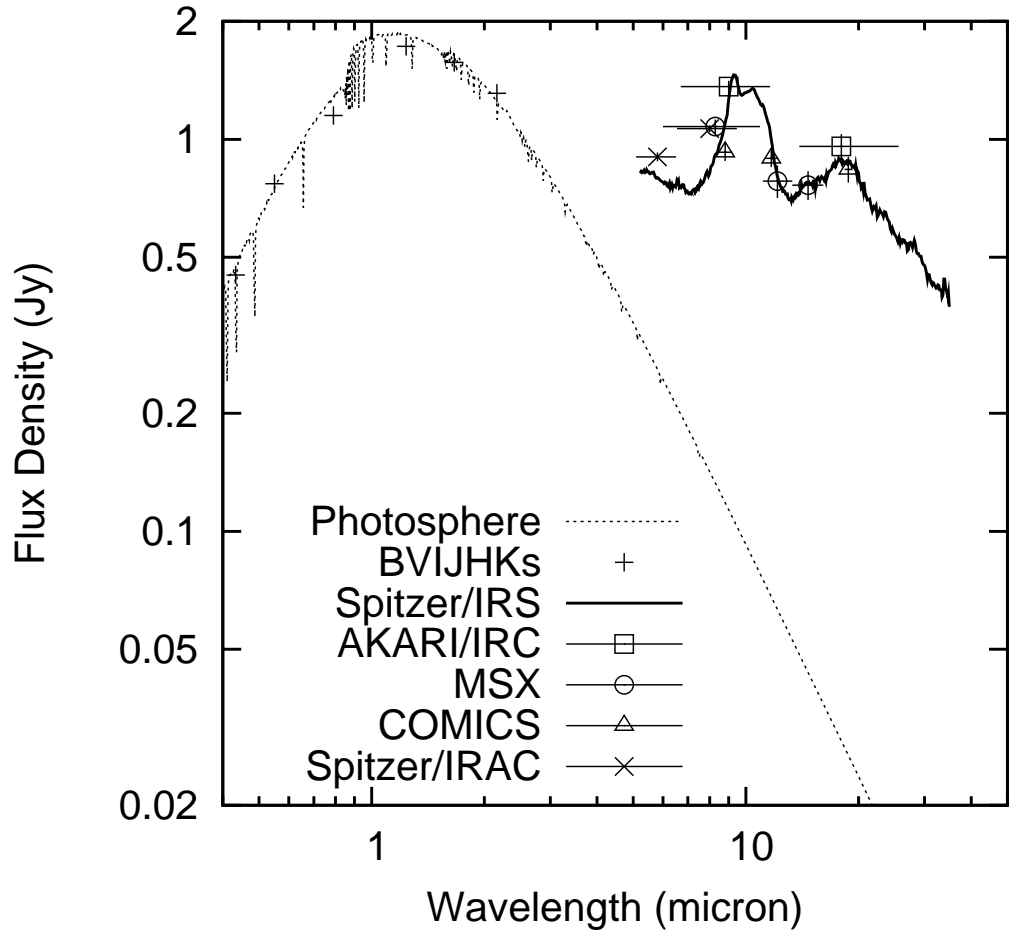}
\caption{
Spectral energy distribution of HD165014. 
The open squares, triangles, circles, crosses, and pluses indicate 
the photometric data obtained with {\it AKARI}/IRC, Subaru/COMICS, {\it MSX}, {\it Spitzer}/IRAC, 
and $BVIJHK_{\rm S}$-band photometry taken from Vizier database, respectively. 
The solid and dotted lines indicate the {\it Spitzer}/IRS spectrum 
and the photospheric contribution of an A0V star \citep{kurucz92}, respectively. 
\label{SED}}
\end{figure}

The obtained {\it Spitzer}/IRS spectrum of the star is shown in Figure~\ref{SED} together with the {\it AKARI} 
and Subaru/COMICS photometry. 
The measurements of the {\it MSX} counterpart (MSX6C G009.0807+00.3009) 
and the {\it Spitzer}/IRAC GLIMPSE II counterpart (SSTGLMC G009.0801+00.3011)
are also shown in Figure~\ref{SED}. 
Most of the MIR flux densities are in good agreement with each other, 
while those of {\it AKARI} 18$~\micron$ band and {\it Spitzer}/IRAC 8$~\micron$-band are slightly higher than the others.
The expected photospheric emission is also plotted in Figure~\ref{SED}. 
Significant excess emission at wavelengths longer than 5$~\micron$ is clearly seen. 
The detected flux densities at 9 and 18$~\micron$ are 10 and 30 times larger than the photosphere, respectively, 
unambiguously indicating the presence of a warm and bright debris disk. 
The slope of the observed spectrum at $\lambda \gtrsim 20~\micron$ 
is consistent with that of the Rayleigh-Jeans side of a blackbody, 
suggesting the dust temperature of $\gtrsim 300$~K 
and the truncation of the disk at the radius where the dust temperature is $\sim 300$~K. 

One of the observable indicators of dust abundance is its fractional luminosity,
a ratio of the infrared luminosity from the disk to the stellar luminosity.
The fractional luminosity of HD165014 is estimated as $\sim 5 \times 10^{-3}$. 
It is comparable to $\beta$~Pic \citep{barrado99},
suggesting that HD165014 is one of the brightest debris disks discovered so far.
The location of the star, deep in the galactic plane,
makes it difficult to find until the {\it AKARI} survey.

\subsection{Features in the Spectrum of the Excess}

To examine the excess emission, we subtract the expected photospheric emission from the observed IRS spectrum. 
The photosphere-subtracted spectrum shows significant emission features centered at around 10 and 20$~\micron$. 
A ubiquitous dust species family, silicate, 
which has features around $\sim 10~\micron$ (due to Si-O stretching modes) 
and $\sim 20~\micron$ (due to O-Si-O bending modes), seems to be a main carrier of the observed features. 
Particularly, the spectrum shows several fine-structured features attributable to 
crystalline silicates. 

\begin{figure}
\epsscale{0.8}
\plotone{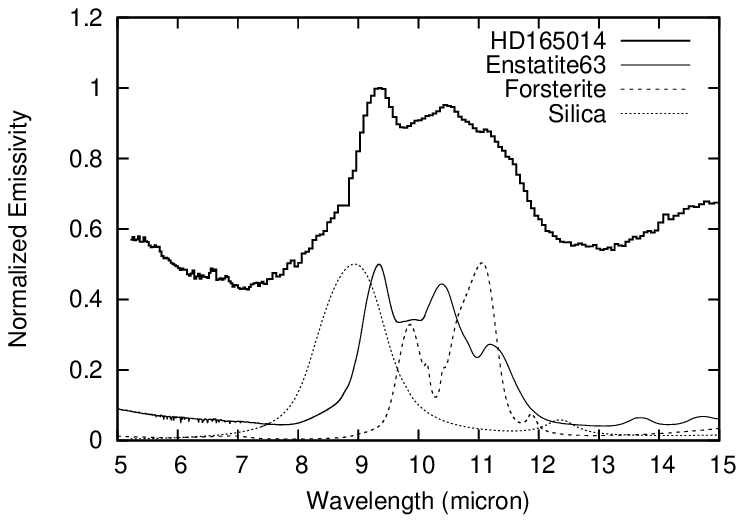}
\plotone{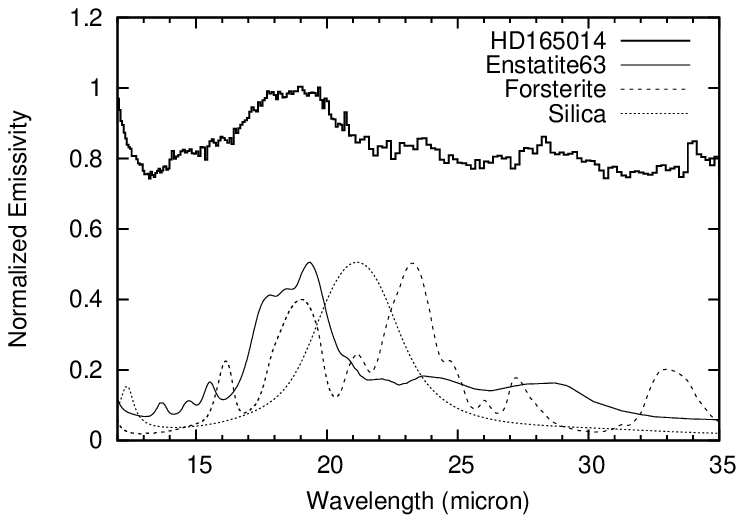}
\caption{Top: emissivity of the excess emission of HD165014 at 5--15$~\micron$ 
derived by division of the observed spectrum by a blackbody of $T=560$~K (thick solid line). 
For comparison, spectra of crystalline enstatite, forsterite, and fused silica 
from laboratory measurements 
are also plotted as thin solid, dashed, and dotted lines, respectively. 
Bottom: emissivity of the excess emission of HD165014 at 12--35$~\micron$
derived by division of the observed spectrum by a blackbody of $T=330$~K. 
The line styles are the same as the top panel.
\label{em}}
\end{figure}

To investigate the details of the features,
Figure~\ref{em} plots the normalized emissivity of the excess emission of HD165014 
by dividing the photosphere-subtracted spectrum by a blackbody ($B_\nu(T)$) of a temperature $T$, 
where $T$ is chosen so that the spectrum level of the feature-free region 
(8.0 and $13.2~\micron$ for the top panel, 
and 13.5 and $33~\micron$ for the bottom panel of Figure~\ref{em}) becomes almost flat. 
The chosen $T$ is 560 and 330~K for the top and bottom panels of Figure~\ref{em}, respectively. 
For comparison, spectra of crystalline enstatite, forsterite \citep{tamanai06}, and fused silica \citep{koike89}
are also plotted in Figure~\ref{em}. 

A significant broad feature is seen at 9--$12~\micron$ in the spectrum. 
Since the width of this feature is large (FWHM $\sim 2.8~\micron$), 
it seems to originate from $\micron$-sized amorphous silicate rather than sub-$\micron$-sized silicate 
\citep{vanboekel03}. 
In addition, some peaks are also seen on the broad feature. 
The most significant peak is located at $9.3~\micron$, and the second at $10.5~\micron$. 
Weaker narrow features are also seen around 11.1--$11.5~\micron$. 
The positions of all the narrow features are consistent with the spectrum of crystalline enstatite. 
The $9.8~\micron$ feature of crystalline forsterite, which is the most common form of crystalline silicate, 
is not seen in the excess emission of HD165014. 
Only weak narrow features are seen around 11.1--$11.5~\micron$, where a strong crystalline forsterite feature 
is expected, in the spectrum of HD165014. 

To analyze the $N$-band spectrum quantitatively and confirm the identification of the features described above, 
we make a simple spectral model fitting of the derived emissivity at 8--$13~\micron$. 
We consider four dust species, crystalline forsterite and enstatite \citep{tamanai06} 
and amorphous olivine and pyroxene \citep{dorschner95}, in our model fit. 
The model spectrum is given by 
\begin{eqnarray}
{\rm Emissivity} = a_{\rm cont}+a_{\rm ol}Q_{\rm ol}+a_{\rm px}Q_{\rm px}+a_{\rm fo}Q_{\rm fo}+a_{\rm en}Q_{\rm en}, 
\end{eqnarray}
where $Q_{\rm ol}, Q_{\rm px}, Q_{\rm fo}$, and $Q_{\rm en}$ are 
the absorption efficiencies of amorphous olivine, amorphous pyroxene, 
crystalline forsterite, and crystalline enstatite, respectively. 
The scaling factors, $a$s, are free parameters ($a_{\rm cont}$ is the contribution of continuum). 
For amorphous olivine and pyroxene, 
we choose one dust size from 0.1, 1.0, 1.5, or 2.0$~\micron$, 
whose absorption coefficients are computed from the optical constants of \cite{dorschner95} 
based the Mie theory \citep{bohren83}. 
We employ a least-squares minimization method to determine the most likely parameters. 
A model with 2.0$~\micron$ sized amorphous silicates is found to provide the best fit 
among all the considered dust sizes of amorphous silicates. 
The derived best-fit parameters are 
$a_{\rm cont}=0.44, a_{\rm ol}=0.00, a_{\rm px}=0.24, a_{\rm fo}=0.03$, and $a_{\rm en}=0.32$. 
The best-fit model spectrum is shown in the top panel of Figure~\ref{bestfit}, 
indicating that the $N$-band spectrum of HD165014 is well reproduced by the model spectrum. 
The broad and narrow features seen in the observed spectrum are indeed attributed to 
$2~\micron$ sized amorphous pyroxene and crystalline enstatite, respectively. 
The contribution of forsterite dust is small and the mass ratio of forsterite to enstatite is estimated as $\sim 1/20$. 

In the longer wavelength region of the spectrum, 
a significant trapezoidal feature is seen at 17--$20~\micron$, 
which is very similar to the enstatite spectrum. 
A broad feature seen around $28~\micron$ also seems to originate from enstatite. 
A weak feature is seen around $23~\micron$, 
which may be accounted for by both forsterite and enstatite. 
The feature around $23~\micron$ might be attributable to forsterite; 
forsterite, if any, is less abundant than enstatite since the feature is very weak.  
Silica features at $9.0$ and 20--$21~\micron$ are not seen in the excess emission of HD165014. 

The observed features and their identifications are summarized in Table~\ref{feature}. 
From the overall shape of the MIR excess spectrum of HD165014, 
we conclude that most of the fine features are attributable to crystalline enstatite 
and that forsterite is much less abundant than enstatite around the star. 
The feature-to-continuum ratios in the 10 and $20~\micron$ regions are $\sim 1$ and $\sim 0.25$, respectively. 
The contribution from the continuum component is large in the excess emission of HD165014, 
suggesting the presence of abundant featureless large dust grains ($>10~\micron$ in size) around the star.

\subsection{Radial Distribution of Dust}

In the top panel of Figure~\ref{em}, 
a rising continuum toward shorter wavelengths $\lambda \lesssim 7~\micron$ is seen, 
suggesting the presence of hotter ($T>560$~K) materials around the star.
Although the rising continuum might also be attributable to amorphous carbon, 
whose emissivity rises toward shorter wavelengths in the NIR and MIR \citep{jager98}, 
no more discussion can be given with the present data. 
The temperature of the hottest component is estimated as $T=750$~K 
from a fitting of the $5-7~\micron$ excess spectrum with a single-temperature blackbody 
assuming that the $5-7~\micron$ excess is coming from the innermost hottest region. 
Therefore dust grains are distributed in the region with temperatures of $T=750-300$~K around HD165014. 
The inner and outer radii of the debris disk are estimated as $0.70$~AU and $4.4$~AU, respectively, 
by assuming that the debris dust is a blackbody particle. 
Possible large circumstellar extinction suggests that the disk is seen nearly edge-on.

\section{Discussion}

As mentioned above, 
it is evident that crystalline enstatite is abundant in the debris disk around HD165014
and forsterite features are not clearly seen. 
\cite{chen06} conduct a comprehensive MIR spectroscopic survey of debris disks. 
Among their 59 debris disk samples, five objects (HR3927, $\eta$~Crv, HD113766, HR7012, and $\eta$~Tel) 
are found to possess spectral features 
that are well-modeled by $\micron$-sized amorphous and crystalline silicates. 
Forsterite features are dominant compared to enstatite toward four samples 
(HR3927, $\eta$~Crv, HD113766, and $\eta$~Tel). 
HR7012, which shows a very strong peak at 9.1--$9.2~\micron$ in its spectrum, 
was initially modeled by enstatite-rich silicate by \cite{chen06}. 
However, \cite{lisse09} conclude that the observed 9.1--$9.2~\micron$ feature 
originates from abundant silica grains 
and that the feature strengths of enstatite and forsterite are almost comparable. 
Although there are a few additional debris disk stars with significant dust features in the $N$-band 
\citep[$\beta$~Pic, HIP8920, HD145263;][]{knacke93,song05,honda04}, 
no sample with high abundance of enstatite is reported to date. 
HD165014 is the only known debris disk that shows strong enstatite features 
compared to forsterite. 

In the solar system, a number of 
achondrite meteorites, which are primarily composed of enstatite, named Aubrites, 
have been discovered. 
E-type asteroids are thought to have enstatite achondrite surfaces and 
to be parent bodies of aubrites \citep[e.g.][]{zellner77} 
based on their reflection spectra. 
E-type asteroids form a large proportion of asteroids inward of the main belt known as Hungaria asteroids 
and are believed to be fragments of differentiated larger bodies that were heated at least to 1700~K \citep{keil89}. 
Considering that enstatite is dominant around HD165014 rather than forsterite, 
the observed debris dust may originate from an analog of E-type asteroids in the solar system, 
being harmonic with the idea that the debris disk is formed by means of collisions between asteroids 
\citep{backman93}. 
It is also known that Mercury's surface is rich in enstatite \citep{sprague98}.
If a Mercury-like planet exists around HD165014 and a mechanism to scatter 
the surface material of the planet, for example infall of small bodies, occurs, 
an enstatite-rich debris disk might be formed. 
It should also be noted that Mercury's high density is interpreted as a result of 
stripping of the surface crustal material \citep{benz88}. 
The possible link between the debris material around HD165014 and Mercury needs to be further explored. 

\begin{figure}
\epsscale{0.7}
\plotone{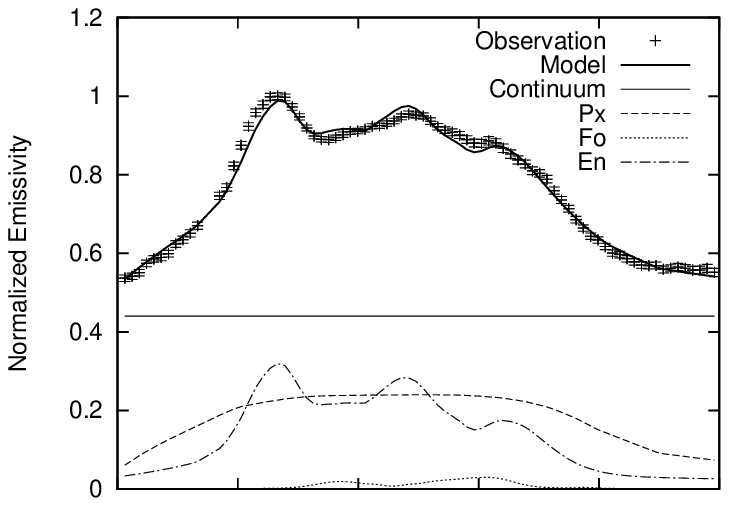}
\vspace{-15pt}
\plotone{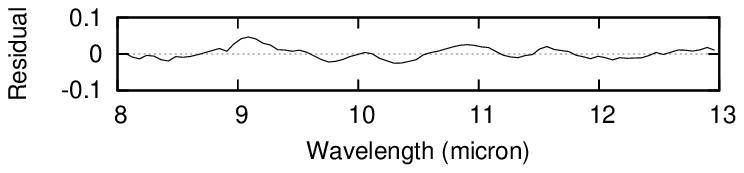}
\plotone{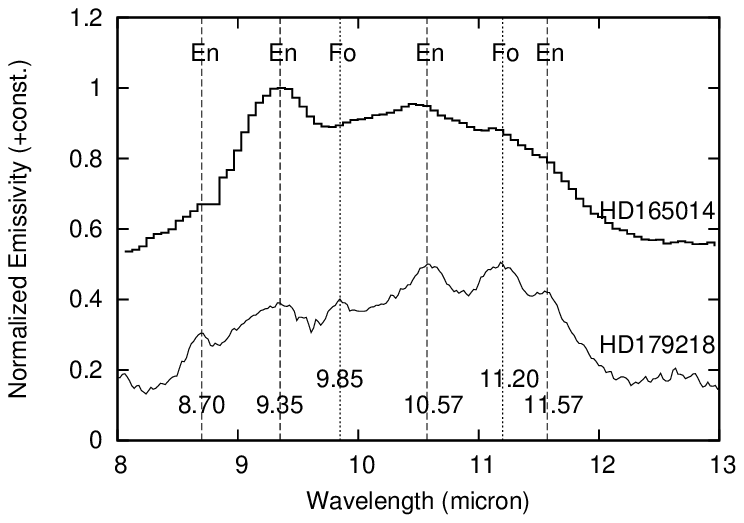}
\caption{Top: fit result of the $N$-band emissivity of HD165014. 
The thick solid line is the best-fit model spectrum, 
which is the sum of constant continuum (thin solid line), 
$2.0~\micron$-sized amorphous pyroxene (dashed line), 
crystalline forsterite (dotted line), and enstatite (dot-dashed line). 
Amorphous olivine component is not shown since no contribution of it is required 
in the best-fit model. 
The residual spectrum subtracted by the best-fit model is shown below the fit result.
Bottom: comparison of the $N$-band spectra of HD165014 (thick solid line) and 
enstatite-rich Herbig Ae/Be star HD179218 \citep[thin solid line;][]{vanboekel05}. 
The vertical dashed and dotted lines indicate peak positions of crystalline enstatite (En) and forsterite (Fo)
detected toward HD179218, respectively.
\label{bestfit}}
\end{figure}

Although a number of MIR spectra of protoplanetary disks associated with younger stars, 
Herbig Ae/Be and T Tau stars, have been collected so far, 
objects that show enstatite features are rare. 
HD179218 is a Herbig Ae/Be star, which has the most enstatite-rich spectrum known to date \citep{bouwman01,schutz05}. 
We show the $N$-band spectrum of HD179218 obtained with TIMMI2 \citep{vanboekel05} in the bottom panel of Figure~\ref{bestfit}. 
While crystalline enstatite features are clearly seen in the spectrum, 
features at 9.85 and $11.20~\micron$ attributable to forsterite are also seen 
with comparable strengths as enstatite, 
suggesting the presence of abundant crystalline forsterite as well as enstatite. 
Annealing experiments of a magnesium silicate smoke made by \cite{rietmeijer86} 
show that the initially formed forsterite and silica react with each other 
and form enstatite by the following reaction
\begin{eqnarray}
{\rm Mg}_2{\rm SiO}_4 + {\rm SiO}_2 \longrightarrow 2{\rm MgSiO}_3.
\end{eqnarray}
\cite{bouwman01} suggest that the presence of enstatite around HD179218 might be 
due to the high luminosity ($300L_\odot$), which gives rise to rapid dispersal of the gas, 
resulting in a high efficiency of the reaction. 
\cite{vanboekel05} suggest that enstatite might be produced by means of chemical equilibrium processes 
in high temperature environments, i.e., inner regions of the protoplanetary disk.  
\cite{sato06} show that crystalline enstatite can be produced 
by simultaneous evaporation of SiO grains and Mg vapor in a plasma  
in laboratory experiments. 
It is now widely accepted that the turbulence in disks is attributed to
the magnetorotational instability \citep[MRI;][]{balbus91}, 
suggesting that at least part of the disk is sufficiently ionized 
\citep[e.g.][]{inutsuka05}. 
Therefore the formation process of crystalline enstatite suggested by \cite{sato06} 
may work efficiently in protoplanetary disks. 
It is possible that 
crystalline enstatite grains produced in the protoplanetary disk were once stored in 
small bodies such as comets and have been released recently in the HD165014 system. 
Indeed crystalline enstatite is detected toward some comets in the solar system 
by astronomical MIR observations \citep[e.g.][]{lisse06} 
as well as by the STARDUST sample return mission \citep[e.g.][]{zolensky06}. 
However, it remains an open question why crystalline forsterite is depleted around HD165014. 
Further experimental and theoretical studies are required 
to discuss the origin of abundant crystalline enstatite in the debris disk around HD165014 in detail.

\acknowledgments

This research is in part based on observations with {\it AKARI}, a JAXA project with the participation of ESA. 
It is also based on observations with {\it Spitzer}, which is operated by the Jet Propulsion Laboratory, 
California Institute of Technology, under a contract with NASA, 
and with Subaru Telescope, which is operated by the National Astronomical Observatory of Japan. 
We thank A.~Tamanai for providing us with silicate spectra, 
R.~van~Boekel for providing us with Herbig Ae/Be spectra, 
and the anonymous referees, M.~Honda, Y.~K.~Okamoto, C.~M.~Lisse, H.~Chihara, Y.~Imai, A.~Takigawa, D.~Kato, 
and H.~Kimura for their useful comments and suggestions. 
This research was supported by the MEXT, ``Development of Extrasolar Planetary Science.'' 
H. F. is supported by the JSPS.

{\it Facilities:} \facility{{\it AKARI} (ISAS/JAXA)}, \facility{{\it Spitzer} (NASA)}, 
\facility{Subaru (NAOJ)}.

%%%%%%%%%%% References

\clearpage

%________________________________________________________________
%% Table: ALL Photometry, F*, and chi
%________________________________________________________________

\begin{table}
\begin{center}
\caption{Infrared Photometry of HD165014.\label{photometry}}
\begin{tabular}{ccccc}
\tableline\tableline
$\lambda$   & $F_\nu$       & Instrument & Photosphere\tablenotemark{a} & Significance, $\chi$\tablenotemark{b}\\
($\micron$) & (Jy)          &            & (Jy)                         & \\
\tableline
5.8         & $0.90 \pm 0.02 $ & {\it Spitzer}/IRAC\tablenotemark{c}       & 0.26  & 27 \\
8.0         & $1.06 \pm 0.02 $ & {\it Spitzer}/IRAC\tablenotemark{c}       & 0.14  & 43 \\
8.28        & $1.08 \pm 0.05 $ & {\it MSX}\tablenotemark{d}       & 0.14  & 22 \\
8.8         & $0.93 \pm 0.09 $ & COMICS          & 0.12  & 10 \\
%9           & $1.36 \pm 0.04 $ & {\it AKARI}/IRC & 0.15  & 30 \\
9           & $1.36 \pm 0.10 $ & {\it AKARI}/IRC & 0.14  & 13 \\ % absolute uncertainty 7%
11.7        & $0.89 \pm 0.09 $ & COMICS          & 0.069  & 10  \\
12.13       & $0.78 \pm 0.07 $ & {\it MSX}\tablenotemark{d}       & 0.064  & 11 \\
14.65       & $0.76 \pm 0.06 $ & {\it MSX}\tablenotemark{d}       & 0.044  & 13 \\
18          & $0.96 \pm 0.14 $ & {\it AKARI}/IRC & 0.030  & 7 \\ % absolute uncertainty 15%
%18          & $0.96 \pm 0.01 $ & {\it AKARI}/IRC & 0.030  & 200 \\
18.8        & $0.84 \pm 0.01 $ & COMICS          & 0.027  & 220 \\
\tableline
\end{tabular}
\tablenotetext{a}{From the Kurucz model to fitted to $BVIJHK_{\rm s}$-bands data.}
\tablenotetext{b}{$\chi = ({\rm Observed} - {\rm Kurucz}) / {\rm noise}$.}
\tablenotetext{c}{The {\it Spitzer} GLIMPSE II counterpart (SSTGLMC G009.0801+00.3011), whose positional offset from HD165014 is $0\farcs10$. See http://www.astro.wisc.edu/glimpse/glimpsedata.html.}
\tablenotetext{d}{The {\it MSX} counterpart (MSX6C G009.0807+00.3009), whose positional offset from HD165014 is $2\farcs58$.}
\end{center}
\end{table}

\clearpage

%________________________________________________________________
%% Table: Feature ID
%________________________________________________________________

\begin{table}
\begin{center}
\caption{Features in Excess Emission of HD165014.\label{feature}}
\begin{tabular}{lll}
\tableline\tableline
$\lambda$   & Comment      & Identification \\
($\micron$) &              &                \\
\tableline
10 (7.3--12.4) & Very strong broad feature           & Silicate \\
9.3         & Strong peak within $10~\micron$ feature & Crystalline En \\
10.5        & Strong peak within $10~\micron$ feature & Crystalline En \\
11.1        & Peak within $10~\micron$ feature        & Crystalline En/Fo \\
11.5        & Shoulder within $10~\micron$ feature    & Crystalline En \\
14.0--15.0 & Weak feature                            & Unidentified \\
18 (17.8--19.7) & Strong trapezoidal emission        & Crystalline En \\
22.8--24.3 & Weak feature                            & Crystalline En/Fo \\
27.3--30.3 & Weak feature                            & Crystalline En \\
33.8--34.5 & Possible weak feature, low signal-to-noise ratio          & Unidentified \\
\tableline
\end{tabular}
\end{center}
\end{table}

\clearpage

%% Figure SED

\clearpage

%% Figure Emissivity 5-15

\clearpage

%% Figure Best-fit Model & Comparison with HD179218

\clearpage


\begin{thebibliography}{}
\bibitem[Backman 
\& Paresce(1993)]{backman93} Backman, D.~E., \& Paresce, F.\ 1993, Protostars and Planets III, ed. E.~H.~Levy \& J.~I.~Lunine
(Tucson, AZ: Univ. Arizona Press), 1253 
\bibitem[Balbus 
\& Hawley(1991)]{balbus91} Balbus, S.~A., \& Hawley, J.~F.\ 1991, \apj, 376, 214 
\bibitem[Benz et al.(1988)]{benz88} Benz, W., Slattery, W.~L., 
\& Cameron, A.~G.~W.\ 1988, Icarus, 74, 516 
\bibitem[Barrado y Navascu{\'e}s et al.(1999)]{barrado99} Barrado 
y Navascu{\'e}s, D., Stauffer, J.~R., Song, I., 
\& Caillault, J.-P.\ 1999, \apjl, 520, L123 
\bibitem[Bohren 
\& Huffman(1983)]{bohren83} Bohren, C.~F., \& Huffman, D.~R.\ 1983, Absorption and Scattering of Light by Small
Particles (New York: Wiley)
\bibitem[Bouwman et 
al.(2001)]{bouwman01} Bouwman, J., Meeus, G., de Koter, A., Hony, S., Dominik, C., \& Waters, L.~B.~F.~M.\ 2001, \aap, 375, 950 
\bibitem[Chen et al.(2006)]{chen06} Chen, C.~H., et al.\ 2006, 
\apjs, 166, 351 
\bibitem[Cohen et al.(1999)]{cohen99} Cohen, M., Walker, R.~G., 
Carter, B., Hammersley, P., Kidger, M., 
\& Noguchi, K.\ 1999, \aj, 117, 1864 
\bibitem[Dorschner et 
al.(1995)]{dorschner95} Dorschner, J., Begemann, B., Henning, T., Jaeger, C., 
\& Mutschke, H.\ 1995, \aap, 300, 503 
\bibitem[Fitzpatrick 
\& Massa(2009)]{fitzpatrick09} Fitzpatrick, E.~L., \& Massa, D.\ 2009, \apj, 699, 1209 
\bibitem[Fujiwara et 
al.(2010)]{fujiwara09c} Fujiwara, H., et al.\ 2010, \aap, submitted
\bibitem[Honda et al.(2004)]{honda04} Honda, M., et al.\ 2004, 
\apjl, 610, L49 
\bibitem[Houk 
\& Cowley(1975)]{houk75} Houk, N., \& Cowley, A.~P.\ 1975, Michigan Catalog of Two-dimensional Spectral Types for HD Stars, 
Vol.\ 1 (Ann Arbor, MI: Univ. Michigan Dept. Astron.)
\bibitem[Houck et al.(2004)]{houck04} Houck, J.~R., et al.\ 
2004, \apjs, 154, 18 
\bibitem[Inutsuka 
\& Sano(2005)]{inutsuka05} Inutsuka, S., \& Sano, T.\ 2005, \apjl, 628, L155 
\bibitem[Ishihara et 
al.(2010)]{ishihara09} Ishihara, D., et al.\ 2010, \aap, in press (arXiv:1003.0270)
\bibitem[Jager et 
al.(1998)]{jager98} Jager, C., Mutschke, H., \& Henning, T.\ 1998, \aap, 332, 291 
\bibitem[Kataza et al.(2000)]{kataza00} Kataza, H., Okamoto, Y., 
Takubo, S., Onaka, T., Sako, S., Nakamura, K., Miyata, T., 
\& Yamashita, T.\ 2000, \procspie, 4008, 1144 
\bibitem[Keil et al.(1989)]{keil89} Keil, K., Ntaflos, T., 
Taylor, G.~J., Brearley, A.~J., \& Newsom, H.~E.\ 1989, \gca, 53, 3291 
\bibitem[Knacke et al.(1993)]{knacke93} Knacke, R.~F., 
Fajardo-Acosta, S.~B., Telesco, C.~M., Hackwell, J.~A., Lynch, D.~K., 
\& Russell, R.~W.\ 1993, \apj, 418, 440 
\bibitem[Koike et al.(1989)]{koike89} Koike, C., Komatuzaki, T., Hasegawa, H., 
\& Asada, N.\ 1989, \mnras, 239, 127 
\bibitem[Kurucz(1992)]{kurucz92} Kurucz, R.~L.\ 1992, in IAU Symp. 149, The 
Stellar Populations of Galaxies, ed. B.~Barbuy \& A.~Renzini (Dordrecht: Kluwer), 225 
\bibitem[Lecavelier Des Etangs et 
al.(1996)]{lecavelier96} Lecavelier Des Etangs, A., Vidal-Madjar, A., \& Ferlet, R.\ 1996, \aap, 307, 542 
\bibitem[Lisse et al.(2009)]{lisse09} Lisse, C.~M., Chen, 
C.~H., Wyatt, M.~C., Morlok, A., Song, I., Bryden, G., 
\& Sheehan, P.\ 2009, \apj, 701, 2019 
\bibitem[Lisse et al.(2006)]{lisse06} Lisse, C.~M., et al.\ 
2006, Science, 313, 635 
\bibitem[Lord(1992)]{lord92} Lord, S.~D.\ 1992, 
A New Software Tool for Computing Earth's Atmospheric 
Transmission of Near- and Far-Infrared Radiation (NASA Tech. Memo 
103957; Washington, DC: NASA) 
\bibitem[Murakami et al.(2007)]{murakami07} Murakami, H., et al.\ 
2007, \pasj, 59, 369 
\bibitem[Onaka et al.(2007)]{onaka07} Onaka, T., et al.\ 2007, 
\pasj, 59, 401 
\bibitem[Rietmeijer et al.(1986)]{rietmeijer86} Rietmeijer, 
F.~J.~M., Nuth, J.~A., \& MacKinnon, I.~D.~R.\ 1986, Icarus, 66, 211 
\bibitem[Sato et 
al.(2006)]{sato06} Sato, T., et al.\ 2006, \planss, 54, 617 
\bibitem[Sch{\"u}tz et 
al.(2005)]{schutz05} Sch{\"u}tz, O., Meeus, G., \& Sterzik, M.~F.\ 2005, \aap, 431, 165 
\bibitem[Song et al.(2005)]{song05} Song, I., Zuckerman, B., 
Weinberger, A.~J., \& Becklin, E.~E.\ 2005, \nat, 436, 363 
\bibitem[Skrutskie et al.(2006)]{skrutskie06} Skrutskie, M.~F., et 
al.\ 2006, \aj, 131, 1163 
\bibitem[Sprague 
\& Roush(1998)]{sprague98} Sprague, A.~L., \& Roush, T.~L.\ 1998, Icarus, 133, 174 
\bibitem[Tamanai et al.(2006)]{tamanai06} Tamanai, A., Mutschke, 
H., Blum, J., \& Meeus, G.\ 2006, \apjl, 648, L147 
\bibitem[Tody(1993)]{tody93} Tody, D.\ 1993, in ASP Conf.\ Ser.\ 52, Astronomical Data 
Analysis Software and Systems II, ed. R.~J.~Hanisch, R.~J.~V.~Brissenden, \& J.~Barnes 
(San Francisco, CA: ASP), 173
\bibitem[van Boekel et 
al.(2005)]{vanboekel05} van Boekel, R., Min, M., Waters, L.~B.~F.~M., de Koter, A., Dominik, C., van den Ancker, M.~E., \& Bouwman, J.\ 2005, \aap, 437, 189 
\bibitem[van Boekel et 
al.(2003)]{vanboekel03} van Boekel, R., Waters, L.~B.~F.~M., Dominik, C., Bouwman, J., de Koter, A., Dullemond, C.~P., \& Paresce, F.\ 2003, \aap, 400, L21 
\bibitem[Werner et al.(2004)]{werner04} Werner, M.~W., et al.\ 2004, \apjs, 154, 1 
\bibitem[Zellner et al.(1977)]{zellner77} Zellner, B., Leake, M., 
Williams, J.~G., \& Morrison, D.\ 1977, \gca, 41, 1759 
\bibitem[Zolensky et al.(2006)]{zolensky06} Zolensky, M.~E., et al.\ 2006, Science, 314, 1735 
\end{thebibliography}
\end{document}